\documentclass[epj]{webofc}
\usepackage[varg]{txfonts}   

\woctitle{MESON2018 - the 15$^\textrm{th}$ International Workshop on Meson Physics}
\begin{document}
	%
	\title{Open charm measurements at CERN SPS energies with the new Vertex Detector of the NA61/SHINE experiment}
	
	\author{Anastasia Merzlaya \inst{1,2}\fnsep\thanks{\email{anastasia.merzlaya@cern.ch}} for the NA61/SHINE Collaboration
	}
	
	\institute{Jagiellonian University, Krakow, Poland
		\and
		Saint Petersburg State University, Saint Petersburg, Russia
	}
	
	\abstract{
		The study of open charm meson production provides an efficient tool for detailed investigations of the properties of hot and dense matter formed in nucleus-nucleus collisions.
		The interpretation of the existing data from the CERN SPS suffers from a lack of knowledge on the total charm production rate.
		To overcome this limitation the heavy-ion programme of the NA61/SHINE experiment at CERN SPS has been expanded to allow for precise measurements of particles with a short lifetime.
		A new Small Acceptance Vertex Detector (SAVD) was designed and constructed to meet the challenges of open charm measurements in nucleus-nucleus collisions.
		SAVD was installed in December 2016 for data taking with Pb+Pb collisions at 150$A$ GeV/c. An exploratory set of collected data allowed to validate the general concept of the $D^0$ mesons detection via its $D^0 \rightarrow \pi^+ + K^-$ decay channel and delivered the first direct observation of open charm at SPS energies. In October and November of 2017 a large statistic data set has been taken for Xe+La at 150$A$, 75$A$, and 40$A$ GeV/c, these data are currently under intense analysis.
		The physics motivation behind the open charm measurements at the SPS energies will be discussed. Moreover, the concept of the SAVD hardware and status of the analysis will be shown.
		Also, the future plans of open charm measurements in the NA61/SHINE experiment related to the upgraded Vertex Detector will be presented.
	}
	\maketitle
	\section{Introduction}
	\label{intro}
	The study of open charm production is a sensitive tool for detailed investigations of the properties of hot and dense matter formed in nucleus-nucleus collisions. 
	In particular, charm mesons are of vivid interest in the context of the phase-transition between confined hadronic matter and the quark-gluon plasma.
	The $\langle c\bar{c}\rangle$ pairs produced in the collisions are converted into open charm mesons and charmonia (J/$\psi$ mesons and it's excited states).
	It was suggested that colour screening in the plasma would reduce and eventually prevent the binding of charm quarks and anti-quarks to produce charmonia, thus suppressing charmonium production in nuclear collisions relative to p+p interactions and providing evidence for deconfinement \cite{MatsuiSatz, Muller, Satz}. 
	However, due to initial state effects in nucleus-nucleus reactions like shadowing, parton energy loss etc. \cite{Petridis}, the overall scaled number of the $\langle c\bar{c}\rangle$ pairs produced in nuclear collisions may be reduced.
	Hence the effect of the medium on J/$\psi$ survival can only be determined by studying the charmonium yield relative to the yield of open charm mesons \cite{Satz}.
	
	The J/$\psi$ mesons have been measured at the top SPS energy (158$A$ GeV/c) by NA38/NA50 and NA60 experiments \cite{NA50}.
	Systematic measurements of open charm production are urgently needed for the interpretation of these results. 
	
	Also such study gives a unique opportunity to test the validity of theoretical models based on perturbative Quantum Chromodynamics and Statistical model approaches for nucleus collisions at SPS energies, which provide very different predictions for charm yields.

	
	\section{The NA61/SHINE experiment for open charm measurements}
	\label{sec-1}
	The SPS Heavy Ion and Neutrino Experiment (NA61/SHINE) \cite{NA61} at CERN was designed for studies of the properties of the onset of deconfinement and search for the critical point of strongly interacting matter by investigating p+p, p+A and A+A collisions at different beam momenta from 13$A$ to 158$A$ GeV/c for ions and upto 400 GeV/c for protons.
	
\begin{figure}
		\centering
		\sidecaption
		\includegraphics[width=0.7\linewidth,clip]{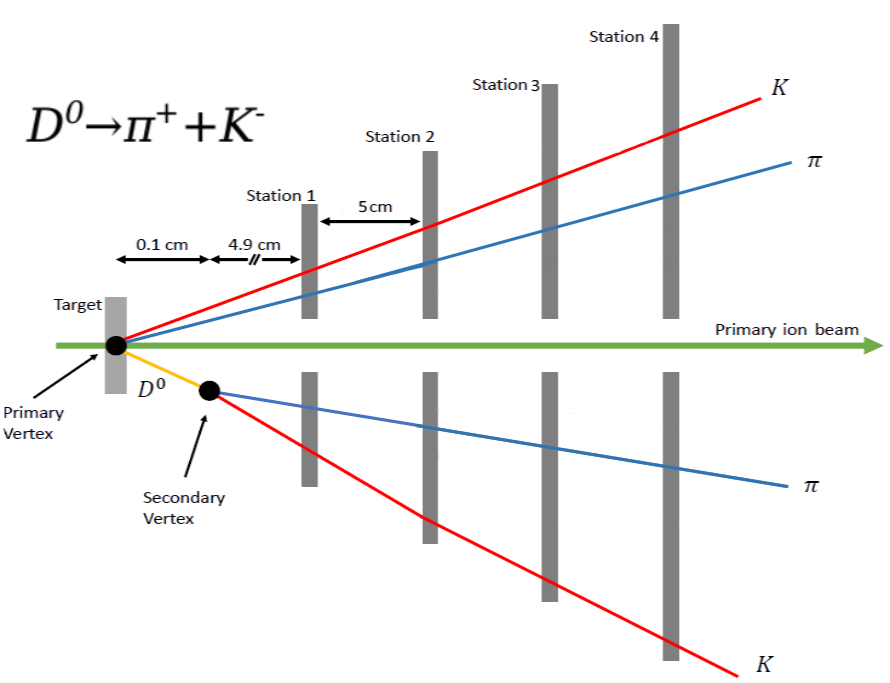}
		\caption{Schematics of reconstruction strategy of $D^0 \rightarrow \pi^+ + K^-$ decay channel with help of Vertex Detector. $\pi$ and $K$, are produced in primary nucleus-nucleus collisions in large numbers by other sources giving huge background. Note, that the vertices are separated not to scale.}
		\label{Figure1}       
	\end{figure}

	To distinguish the daughter particles of $D^0$ mesons from hadrons produced in primary nucleus-nucleus interaction point, one aims to select only hadron pairs generated in a secondary decay vertex (see Fig.~\ref{Figure1}).
	This requires precise reconstruction of the primary and secondary vertices with  an accuracy at the level of tens of microns. 
	This is done by extrapolating the trajectories back to the target and identifying intersection points. 
	The primary vertex will typically appear as intersection point of multiple tracks while the tracks originating from selected decays will intersect at the well-defined displaced point (secondary vertex).
	
	To meet challenges of open charm measurements  the NA61/SHINE experiment was upgraded with the Small Acceptance Vertex Detector (SAVD).  
	The SAVD is placed 5 cm downstream from the target.
	High coordinate resolution MIMOSA-26 sensors \cite{mimosa} based on the CMOS Pixel Sensor technology were chosen as the basic detection element of the SAVD stations.
	The sensors provide a spatial resolution of 3.5 $\mu$m, have very low material budget (50 $\mu$m thickness), and their readout time is 115.2 $\mu$s. The stations are kept in He filled box to reduce beam-gas interactions.
	The simulations of the performance of the SAVD using AMPT event generator as input have shown that about 5\% of all  $D^0 \rightarrow \pi^+ + K^-$ decays in Pb+Pb collisions at 150$A$ GeV/c can be reconstructed in SAVD \cite{sim, Merzlaya2017}.
	
	\section{Open charm measurements with SAVD}
	\label{sec-2}
\begin{figure}
		\centering
		\sidecaption
		\includegraphics[width=0.6\linewidth,clip]{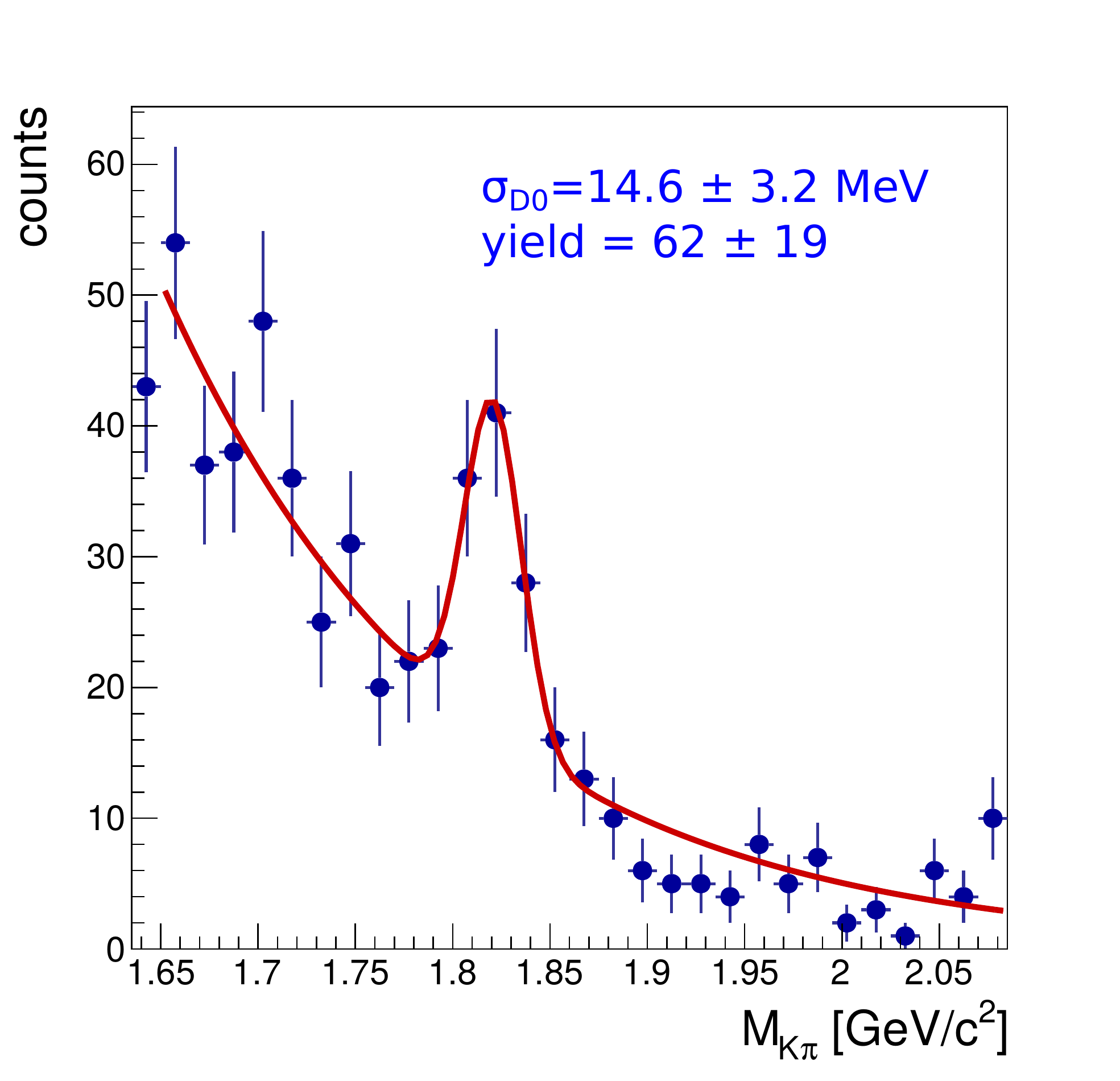}
		\caption{Invariant mass distribution of unlike charge $\pi, K$ decay track candidates for Pb+Pb collisions at 150$A$ GeV/c.}
		\label{Figure2}       
	\end{figure}

	The SAVD was installed as a part of the NA61/SHINE facility and used in 2016 during a Pb+Pb test run at 150$A$ GeV/c beam momenta. An exploratory set of data was collected and analyzed \cite{Merzlaya2018}. 
	The main goal of the test was to prove precise tracking in the large track multiplicity environment, demonstrate the ability of precise vertex reconstruction and extract the physics result. 
	The obtained primary vertex resolution is $\sigma_x$ = 5 $\mu$m, $\sigma_y$ = 1.5 $\mu$m and $\sigma_z$ = 30 $\mu$m (the difference between $\sigma_x$ and $\sigma_y$ is caused by the presence of a vertical component of the magnetic field in the SAVD volume).
	The quality of collected data was sufficient enough to extract a first signal in the $D^0 \rightarrow \pi^+ + K^-$ decay channel (see Fig.~\ref{Figure2}).
	This was the first, direct observation of open charm in nucleus-nucleus collisions at SPS energies.
	
	Successful performance of the SAVD in 2016 led to the decision to use the device during the Xe+La data taking in 2017.
	A large statistic data set has been taken for Xe+La at the beam momenta of 150$A$, 75$A$ and 40$A$ GeV/c.
	During these measurements the thresholds of the SAVD sensors where tuned to obtain high hit detection efficiency.
	This led to significant improvement of vertex resolution precision: $\sigma_x$ = 1.3 $\mu$m, $\sigma_y$ = 1.0 $\mu$m and $\sigma_z$ = 15 $\mu$m.
	The NA61/SHINE experiment is planning high statistics data taking of Pb+Pb collisions at 150$A$ GeV/c in November-December 2018 (the aim is to collect 10 million central collisions).
	This will allow for starting the detailed research programme of the open charm measurements. 
	
	Looking forward, the NA61/SHINE experiment will be upgraded during CERN long-shutdown 2019-2020 to increase the data taking rate from 80Hz to 1kHz \cite{NA61proposal}.
	The upgraded VD will be based on ALPIDE sensors developed for ALICE ITS \cite{alpide}, and will have larger acceptance for each station.
	The proposed programme will allow to perform systematic studies of $D^0$, $\overline{D}^0$, $D^+$ and $D^-$ production. 
	This study will provide $\langle c\bar{c}\rangle$ in central Pb+Pb collisions needed to investigate the mechanism of charm production in this reaction.
	Moreover, the data will allow to establish the centrality dependence of $\langle c\bar{c}\rangle$ in Pb+Pb collisions at 150$A$ GeVc and thus address the question of how the formation of QGP impacts $ \text{J}/\psi $ production.  
	In total, 500M minimum bias Pb+Pb events are expetced to be collected at top SPS energy.
	From this, it is expected that 76k $D^0$ and $\overline{D}^0$ will be collected, of which 31k from most central events.
	Additionally, 46k $D^+$ and $D^-$ will be collected, of which 19k from most central events.

	



	\begin{acknowledgement}
		The work  related to the SAVD hardware development and  experimental data taking was supported by the Polish National Center for Science grant 2014\slash15\slash B\slash ST2\slash02537 and the work related to reconstruction and data analysis was supported by the Russian Science Foundation research grant 16-12-10176.
	\end{acknowledgement}
	%
	%
	%

\end{document}